\documentclass[prl,aps,twocolumn,showpacs]{revtex4}

\usepackage[final]{graphicx}

\begin{document}


\title{First-principles Simulations of the stretching and 
final breaking of Al nanowires:
Mechanical properties and electrical conductance} 


\author{Pavel Jel\'{\i}nek$^{1,2}$ \email{pavel.jelinek@uam.es}, 
Rub\'en P\'erez$^{1}$, Jos\'e Ortega$^{1}$ and Fernando Flores$^{1}$}

\affiliation{
$^{1}$ Departamento de F\'{\i}sica Te\'orica de la Materia Condensada,
Universidad Aut\'onoma de Madrid, E-28049 Spain} 

\affiliation{
$^2$ Institute of Physics,
Academy of Sciences of the  Czech Republic,
Cukrovarnick\'a 10, 1862 53,Prague, Czech Republic}


\date{\today}

\begin{abstract}
The evolution of the structure and conductance of an Al nanowire subject to a 
tensile stress has been studied by first-principles total-energy simulations. 
Our calculations show the correlation between discontinuous changes in the 
force (associated to changes in the bonding structure of the nanowire) and 
abrupt modifications of the conductance
as the nanowire develops a thinner neck, in agreement with the experiments. 
We reproduce the characteristic increase of the conductance in the
last plateau, reaching a value close to the conductance quantum 
$G_0 = 2 e^2 / h$ before the breaking of the nanowire. 
A dimer defines the contact geometry at these last stages, with three
channels (one dominant) contributing to  the conductance.
\end{abstract}


\maketitle

The electrical and mechanical properties of metallic nanowires have received 
a lot of attention \cite{Agrait}. Although point contacts have been studied 
for many years, only recently the gentle control of the distance between two 
metals using an STM-AFM \cite{Rubio} or a mechanically controlled breaking 
junction (MCBJ) \cite{Scheer1} has allowed the experimental 
characterization of atomic 
contacts and the observation of  quantum effects in both the conductance and the 
forces \cite{Rubio}. 
In a~pioneering work, Scheer et al.\cite{Scheer2} have shown, 
analyzing the superconducting properties of an~atomic contact, 
how the transport properties of the system just before the breaking 
point depend on a~few channels that they related to the atomic orbitals 
of the contact.

The formation of  necks and atomic contacts in stretched metallic nanowires 
has been analyzed theoretically using different approaches. 
The nanowire deformation can be studied by molecular
dynamics simulations using either
classical \cite{Landman,Todorov} or effective-medium 
theory potentials \cite{Bradkovsky,Sorensen}.
First principles calculations based on Density Functional theory (DFT)
\cite{Lang,Kobayashi}
provide a more accurate description of the mechanical properties and the
electronic structure needed for the calculation of the conductance, but the
large computational demand restricts most of the applications to the
analysis of model geometries for the contact (e.g. monoatomic chains). 
Up to our knowledge, the most complete calculation of
the deformation of a~metallic nanowire has been presented by Nakamura 
et al \cite{Nakamura}, who analyzed, using DFT calculations, a Na nanowire with   
39 atoms. In this simulation, the wire is elongated  in steps of 0.2 or 0.4 
 \AA\ until it reaches  the breaking point. The conductance is
determined using the Landauer-Buttiker formula, where the transmission matrix is
calculated from the self-consistent electrostatic potential using
scattering techniques \cite{Hirose95}.
This calculation showed how 
the nanowire deformation was accompanied by a rearrangement of the atomic 
configuration, that introduces also jumps in the forces and the 
conductance of the system. 


Addressing this complex problem with a fully-converged first-principles 
description would be still too computationally demanding. One possibility
is to stick to accurate plane-wave (PW) DFT methods, 
carefully relaxing the conditions
for convergence \cite{Kruger02} (basis set cutoff, k-point sampling, etc).
A second alternative is to resort to local orbital MD-DFT methods,
specially those devised with the aim of computational efficiency, that allow
first-principles studies of much more complex systems.
The formulation in terms of local orbitals has an added value, as the transport
properties can be easily calculated from the resulting  (tight-binding or
LCAO) electronic hamiltonian using non-equilibrium Green's function
techniques \cite{Mingo}. 
Thus, efficient local-orbital MD-DFT methods are probably the best
available tools for a first-principles analysis of complex nanowires.
In this work, we have studied the evolution of the structure and the
conductance of an Al nanowire as a function
of the stretching by means of a fast local-orbital minimal basis 
MD-DFT technique (Fireball96)\cite{Demkov}. 
This technique offers a very favorable accuracy/efficiency
balance if the atomic-like basis set is chosen carefully
\cite{Al-Fireball96}.
This approach is complemented with
PW-DFT calculations (using CASTEP\cite{castep}) performed at
the critical points of the nanowire deformation, as discussed below. 
Closely related approaches \cite{Taraschi98,Guo02,Brandbyge,Palacios},
based on the combination of ab-initio calculations on local-orbital basis
and Keldysh-Green function methods, have been recently applied to
characterize some ideal geometries (mainly, atomic chains) for Al and Au
nanowires.

\begin{figure*}
\includegraphics[width=140mm, height=70mm]{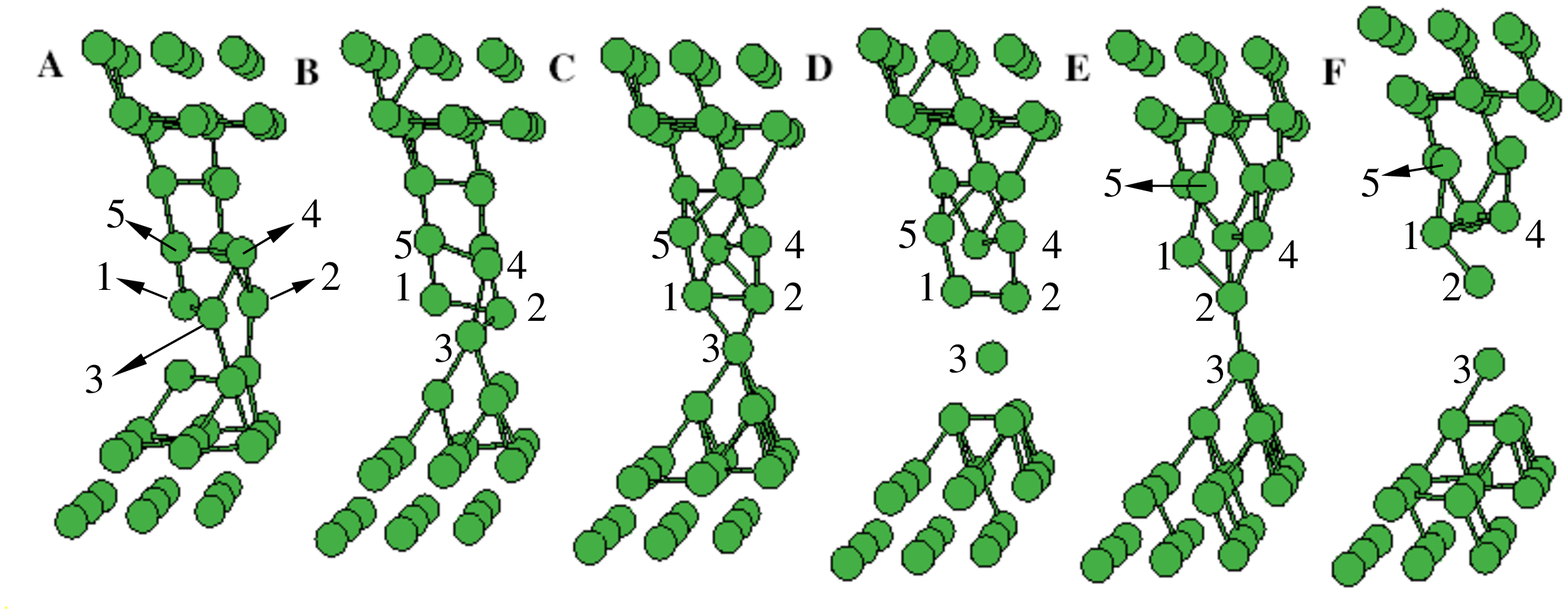}
\caption{Ball-and-Stick model of the structure of the Al nanowire for 
different steps of the stretching process (see fig. \ref{fig:etot}). 
The atoms involved in the important bonding rearrangements related to 
discontinuous changes in total energy, force and
conductance are labelled 1 to 5.}
\label{fig:movie} 
\end{figure*}

\begin{figure}
\hspace*{-0.5cm}\includegraphics[width=70mm,angle=-90]{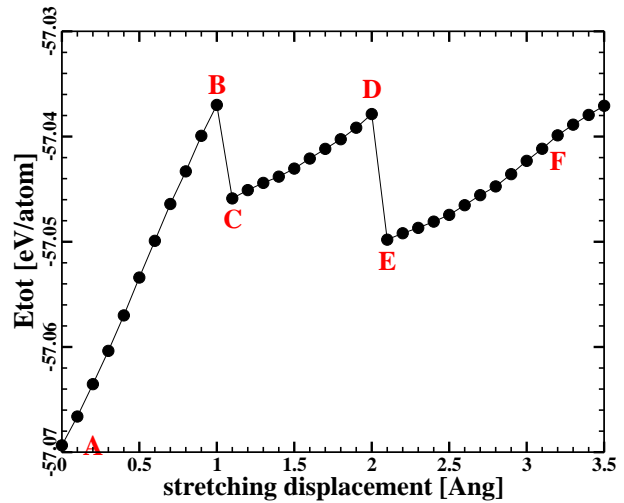}
\caption{Total energy of the Al nanowire as a~function of the stretching 
displacement, $\Delta d$.}
\label{fig:etot}
\end{figure}

We have studied the system shown in figure~\ref{fig:movie}A, 
having 48 atoms --12 of them in the wire--, 
embedded in an~Al(111) surface having a~(3x3) periodicity. 
We also impose periodic conditions in the direction perpendicular to the surface, 
joining artificially the last two layers of the system.
The stretching of the system is simulated increasing the distance between the two 
limiting layers by steps of 0.1 \AA. After each step, the system is allowed
to relax toward its configuration of minimum energy; in this relaxation, 
only the atoms located in the last two layers remain fixed, 
meaning that 30 atoms are allowed to relax.
We have checked the validity
of the calculation performed with Fireball96, repeating similar calculations
with CASTEP at points of the deformation where rearrangement of the atoms appeared 
(see the discussion below).
Figure~\ref{fig:movie} shows snapshots of the nanowire geometry along the 
stretching path; 
different profiles correspond to the points labelled A to F 
in fig.~\ref{fig:etot}, where the 
total energy per atom is shown as~function of the stretching displacement. 
This figure ~\ref{fig:etot}
displays the points where the nanowire suffers an~important rearrangement in its 
geometry: these jumps occur between points B and C, and between points D and E.
Fig ~\ref{fig:movie} shows the initial geometry (A), 
the geometries before and after the first (B,C) and second (D,E) jumps, and
the structure close to the final breaking point (F).
It is important to notice that the main rearrangement of atoms occurs in the 
layer
formed by atoms 1,2 and 3. In particular, along the path AB, the atom 3 starts
to move out of the initial 1-2-3 layer; at the first jump, this atom takes 
an~independent position linking a~lower layer and the two atoms 1-2.
Along the path CD, the geometry of the point C does not change too much, the 
atom 2 moving slowly toward atom 3; eventually, at the second jump, atom 2 
forms a~dimer with atom 3, and the nanowire reaches a~geometry very similar 
to the one formed at the breaking point. In particular, along the EF path 
the system evolves basically changing mostly the distance between atoms 2 and 3.

We have checked these calculations, obtained using a~Fireball code, by recalculating
geometries around the jumps (BC and DE) using CASTEP: these calculations confirm 
the validity of the Fireball results; we have found, however, 
some minor differences.
For instance, in CASTEP we find the system to present the first jump for 
$\Delta d=$ 0.9 \AA, namely, a~step before the jump found in Fireball; 
the system evolves, however, 
across the first jump to the same geometry calculated in Fireball.

\begin{figure}
\hspace*{-0.5cm}\includegraphics[width=70mm,angle=-90]{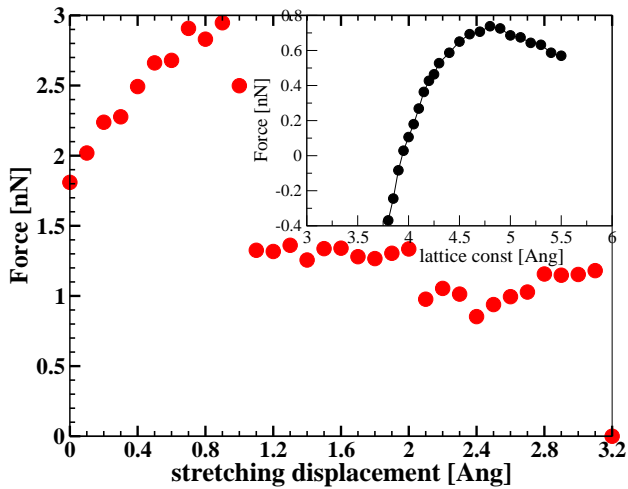}
\caption{Normal force along the stretching path. The inset shows the force 
between n.n. atoms in Al crystal defined as the derivative of the energy 
per atom with respect to the n.n. distance during hydrostatic strain, 
divided by 6 bonds assumed in the f.c.c. structure.}
\label{fig:force}
\end{figure}

Figure~\ref{fig:force} shows the force along the nanowire stretching.  
It is very satisfactory
that the first jump seems to appear for a~maximum in the force; along the path CD 
the force is practically constant and along the final deformation, 
path EF, the force 
increases except for some small fluctuations, until reaching a~maximum at the point 
where the wire breaks. It is interesting to compare the forces shown in 
~\ref{fig:force}, 
with the one calculated at the break point for an Al-crystal: this is defined by 
the maximum derivative of the energy per atom with respect to the 
nearest-neighbor distance for an~Al-crystal,
divided by the number of bonds each atom is assumed to form
(6 bonds per atom in a f.c.c. structure). Our calculations yield 0.7 nN  
for the maximum force, and this quantity should be compared with the forces of 
fig ~\ref{fig:force} that are typically larger than that value: 
at the nanowire break point the force is around 
1.2 nN, 1.7 times the bulk value.

We have calculated the electrical conductance of the nanowire using a~Keldysh-Green 
function approach based on the first-principles 
tight-binding Hamiltonian obtained from the Fireball
code, at each point of the deformation path. In this formalism, we rewrite
this hamiltonian describing the 
system as $\hat H_1 + \hat H_2 + \hat T_{12}$, where the total 
system is splitted into the two parts, 1 and 2, $\hat T_{12}$ defining the 
coupling between both. Typically, we use the thinnest part of the nanowire
to define the interface between these two subsystems. 
Then, the differential conductance \cite{Mingo} is given by:     
   
\begin{eqnarray}
\label{cur}
& & G = \frac{dI}{dV} = \\ \nonumber 
& & = \frac{4\pi e}{h}  
\mbox{\rm Im} Tr 
\bigl[ \hat T_{12} \hat \rho_{22}(E_F) \hat D_{22}^r (E_F) \hat T_{12} 
\hat \rho_{11} (E_F) \hat D_{11}^a(E_F) \bigr] 
\end{eqnarray}  

where $\hat \rho_{11}$ and $\hat \rho_{22}$ are the density matrices associated 
with sides 1 and 2, respectively, while:

\begin{eqnarray}
\label{denom}
\hat D_{22}^r = [\hat 1 - \hat T_{12} \hat g_{22}^r(E_F)\hat T_{21} 
\hat g_{11}^r(E_F) ]^{-1}
\end{eqnarray}
and $\hat D_{11}^a$ is given by a~similar eqn.; $\hat g_{11}^r$ and  
$\hat g_{22}^r$ are the retarded Green function of the decoupled sides,2 and 1. 

\begin{figure}
\hspace*{-0.5cm}\includegraphics[width=70mm,angle=-90]{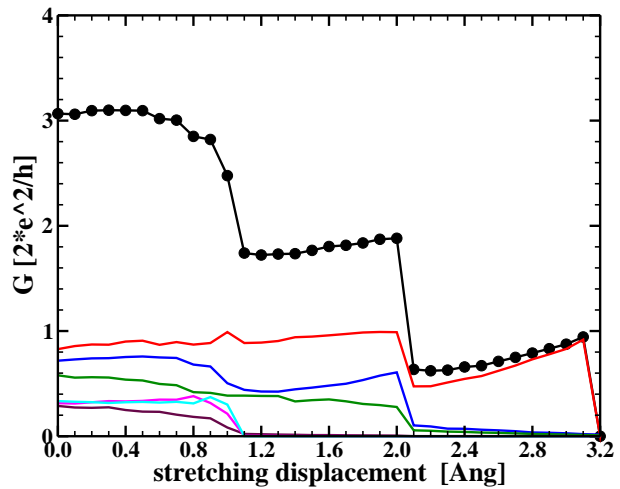}
\caption{Total differential conductance (in units of the conductance quantum) 
and channel contribution along the stretching path.}
\label{fig:channels}
\end{figure}

Using eqn.~\ref{cur}, we calculate the differential conductance along the 
stretching deformation (fig.~\ref{fig:channels}). 
>From this figure we 
see how this conductance changes dramatically at 
the points B-C and D-E: in both cases  the 
conductance jump coincides with the force jump (fig ~\ref{fig:force}), 
and is determined by the atomic rearrangement the system suffers at those points. 
The conductance behavior shows two other important features: (i) the 
nanowire breaks at a~point where the conductance is very close to 
the quantum unit, $\rm G \approx 0.95 \biggl( \frac{2e^2}{h} \biggr)$; (ii)
second, before the breaking point, the conductance takes values a~little 
smaller than the last one, with a~negative slope. These two features are 
in good agreement with the experimental evidence \cite{Cuevas,Agrait}
. Morever, at the breaking point
the system shows a~geometry slightly different to the one 
suggested in \cite{Scheer2}.
In particular, we find that an~Al-dimer is responsible for the 
conductance properties the nanowire shows before it breaks.

Finally, we have also addressed the question of how many different channels 
contribute to the conductance. This has also been analyzed using eqn \ref{cur}; 
this eqn, can be rewritten, using the cyclic property of the trace in the form:

\begin{eqnarray}
\label{cur1}
G = \frac{dI}{dV} =  \frac{4 \pi e}{h} \mbox{\rm Im} Tr(\hat t \hat t^+),
\end{eqnarray}  
where $\hat t = \hat \rho_{11}^{\frac{1}{2}}\hat D_{11}^a \hat T_{12} 
\hat \rho_{22}^{\frac{1}{2}}$ \cite{Buttiker}. 
By diagonalizing this transfer matrix, we find the channels contributing to G.

Figure~\ref{fig:channels} also shows how different channels contribute 
to the total differential
conductance. Near the break point, we find that only three channels yield 
appreciable contributions to the total current,  although there appears 
a~predominant mode,
very much in agreement with the results published by Scheer et al \cite{Scheer2}. 
Along the trajectory CD, before the last jump, the system also presents
three predominant channels, but, in this case, all the three yield 
important contributions to the current; notice that in this case, the total
conductance is around 2 $\bigl( \frac{2e^2}{h} \bigr)$, with a~channel 
contributing practically $\bigl( \frac{2e^2}{h} \bigr)$, and two other 
channels, each one contributing $ \frac{1}{2} \bigl( \frac{2e^2}{h} \bigr)$.
In the initial path, AB in fig 1, 
we find six channels contributing with a~total
conductance of around $3 \bigl( \frac{2e^2}{h} \bigr)$. From this analysis 
we conclude that the channels are not either open or closed; their particular 
contribution depends very much on the geometry of the nanowire contact.
We can only say that channels tend to close along the nanowire deformation, 
and that once a~channel is closed by the stretching deformation it remains 
closed all the time.

In conclusion, we have presented total energy DFT calculations for the 
stretching deformation of an~Al-nanowire. The use of a~fast local-orbital
code has allowed us to analyze in detail the nanowire geometrical 
configurations along the stretching path. At the same time, the first-principles 
tight-binding
Hamiltonian obtained at each geometry  for the electronic system, has allowed 
us to analyze also in a~very efficient way, using Keldish-Green function
methods, the differential conductance of the system. 
Our results reproduce most of the experimental evidence, and show that
the deformation induces at particular points sudden rearrangement of atoms that 
manifest itself in jumps of the forces and the conductance. We also find that in 
the final process of deformation, the nanowire breaks at a~point having 
a~differential conductance close to $\biggl( \frac{2e^2}{h} \biggr)$, and that 
the total conductance is the result of the contribution of three channels, 
one of them being the predominant one.

\begin{acknowledgments}
P.J. gratefully acknowledges financial support by the European project no.
HPRN-CT-2000-00154. This work has been supported by the DGI-MCyT (Spain) 
under contract MAT2001-0665, and by Consejeria de Educaci\'on de la
Comunidad de Madrid under project 07N/0050/2001.
\end{acknowledgments}


\newpage 


\begin{thebibliography}{19}
\bibitem{Al-Fireball96}
The valence electrons were described by $s$ and $p$ slightly excited 
pseudoatomic orbitals \cite{Sankey}. For the cutoff radii $\rm R_c =6.4 a.u.$ 
we obtained for f.c.c. Al a lattice parameter of a = 3.97 \AA\ and a bulk
modulus of B = 74GPa (experiment: a = 4.05 \AA, B = 76 GPa).
\end{thebibliography}
\end{document}